\documentclass[preprint,12pt,3p]{elsarticle}

\usepackage{amssymb}
\usepackage{amsmath}
\usepackage{dcolumn}
\usepackage{bm}

\usepackage{color}
\usepackage{multirow}

\usepackage[utf8]{inputenc}
\usepackage[T1]{fontenc}
\usepackage{caption}
\usepackage{subcaption}
\usepackage{etoolbox}
\usepackage{siunitx}
\usepackage{booktabs}
\usepackage{tabularray}
\UseTblrLibrary{booktabs}
\usepackage{float}
\usepackage[capitalize, noabbrev]{cleveref}

\journal{Materials Today Physics}

\begin{document}

\begin{frontmatter}



\title{From plasma to pattern: observation and characterization of periodic structure formation in dielectric breakdown channels of electron-irradiated insulators}


\author[a]{Nick R. Schwartz\fnref{1}}
\author[a]{Bryson C. Clifford}
\author[a]{Carolyn Chun}
\author[a]{Emily H. Frashure}
\author[a]{Kathryn M. Sturge}
\author[a]{Noah Hoppis}
\author[a]{Holly Wilson}
\author[a]{Meryl Wiratmo}
\author[a]{Jack R. FitzGibbon}
\author[a]{Ethan T. Basinger}
\author[b]{Brian L. Beaudoin}
\author[a]{Raymond J. Phaneuf}
\author[a]{John Cumings}
\author[a]{Timothy W. Koeth\corref{cor1}}

\ead{koeth@umd.edu}
\fntext[1]{Currently at Plasma Science and Fusion Center, Massachusetts Institute of Technology, Cambridge, MA 02139, USA}
\cortext[cor1]{Corresponding author}

\affiliation[a]{organization={Department of Materials Science and Engineering},
		addressline={University of Maryland},
		city={College Park},
		state={MD},
		postcode={20742},
		country={USA}}
\affiliation[b]{organization={Institute for Research in Electronics and Applied Physics},
		addressline={University of Maryland},
		city={College Park},
		state={MD},
		postcode={20742},
		country={USA}}


\begin{abstract}
Dielectric breakdown of insulators is one of the most common failure modes of electronics in the high-radiation environment of space, but its mechanics remain poorly understood. When electron-irradiated polymethyl methacrylate (PMMA) undergoes breakdown, the resulting channels exhibit striking periodic structures with characteristic wavelengths $\sim$80 \textmu m in the recently identified ivy-mode channels. These previously unobserved modulations offer unique insights into the physics of ultra-fast dielectric breakdown. Through materials characterization and theoretical modeling, we identify the physical instability mechanism responsible for these structures. Raman spectroscopy reveals that carbon deposition correlates with channel width variations, indicating that periodic structure formation occurs during the plasma discharge phase. We evaluated three candidate instability mechanisms: the Asaro-Tiller-Grinfeld instability, the Plateau-Rayleigh instability, and the z-pinch entropy mode. The first two mechanisms operate on incompatible timescales and require unphysical material parameters to match observations. In contrast, the z-pinch entropy mode operates during the nanosecond discharge phase and produces wavelengths consistent with plasma densities of 0.1--1\% of solid density and temperatures of 10--100 eV. Current measurements from isolated discharge channels ($\sim$200 A) validate theoretical predictions for the entropy mode. These findings establish that the entropy mode plasma instability during the discharge phase, rather than post-discharge thermal or mechanical processes, govern periodic structure formation in breakdown channels. This work provides new insights into the physics of dielectric breakdown and establishes a framework for predicting discharge morphology and characteristics in insulators.
\end{abstract}

\begin{graphicalabstract}
\includegraphics[width=\textwidth]{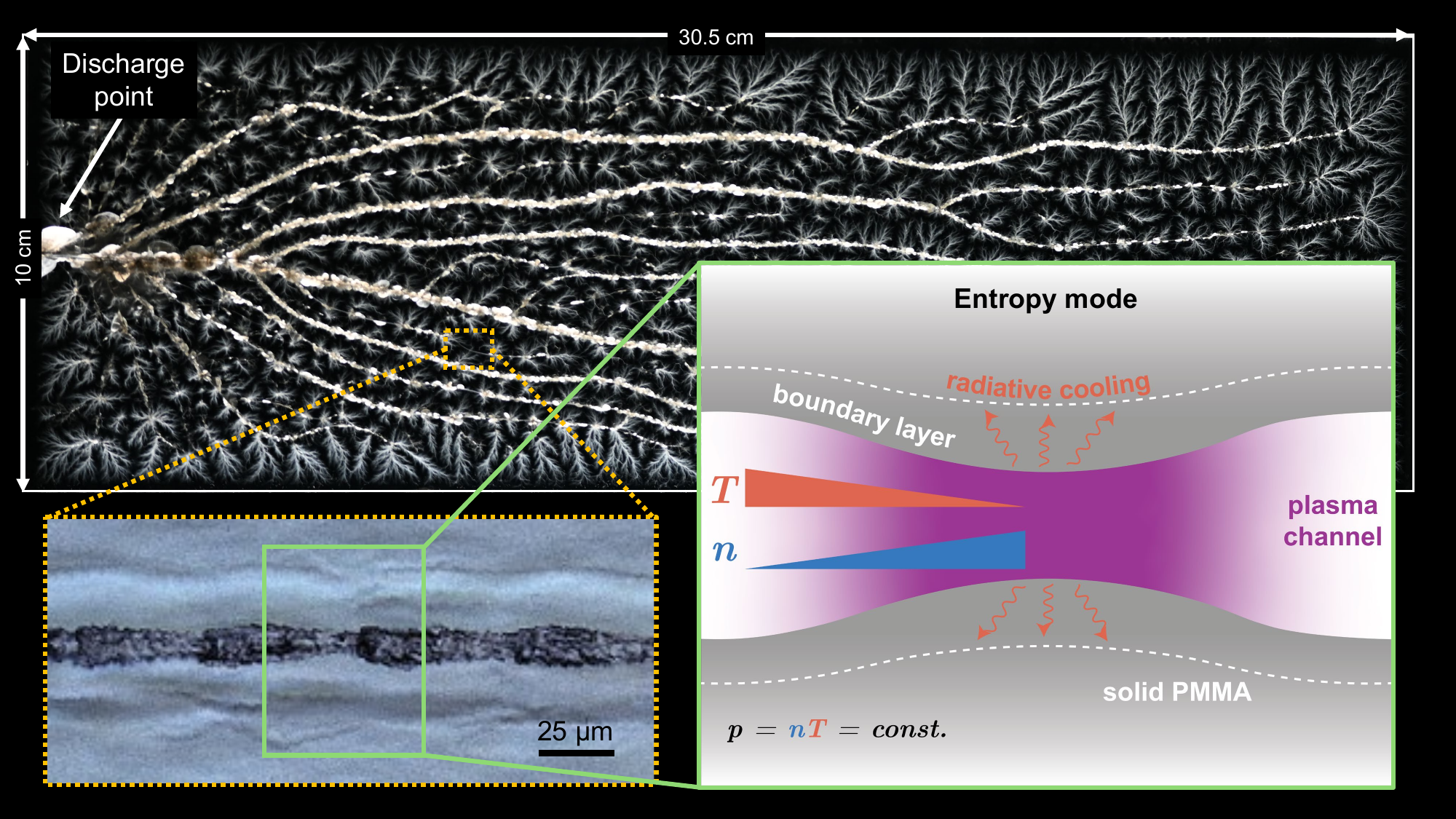}
\end{graphicalabstract}

\begin{highlights}
\item Periodic structures observed in electron-irradiated PMMA breakdown channels
\item Raman spectroscopy shows carbon deposition correlates with channel width
\item Instability identified as z-pinch entropy mode, validated by current measurements
\end{highlights}

\begin{keyword}
dielectric \sep instability \sep breakdown \sep plasma \sep insulator


\end{keyword}

\end{frontmatter}

\section{Introduction}
When insulating materials undergo dielectric breakdown discharge (DB) events, they form distinctive tree-like damage patterns known as Lichtenberg figures (LFs) that propagate through the material at remarkable speeds -- in some cases approaching one-twentieth the speed of light. These breakdown events represent a fundamental challenge for dielectric materials used in sensitive electronics in space, where they are believed to account for over half of satellite failures \citep{cooke1986space,koons1999impact}. The breakdown channels effectively embed conductive pathways within the insulator, permanently altering its dielectric properties, potentially causing mechanical degradation, and generating damaging currents during the discharge process. Understanding the physical mechanisms driving these ultra-fast discharge processes is critical for developing radiation-hard materials and predicting failure modes in harsh environments.

Recent advances in high-speed imaging techniques have enabled the observation of LF formation with nanosecond temporal resolution \citep{Hoppis2023}, revealing previously unknown discharge channel morphologies and propagation dynamics. In electron-irradiated polymethyl methacrylate (PMMA) -- a model dielectric material chosen for its optical transparency and similarity to spacecraft insulators -- a variety of different types of electrical trees have been observed \citep{Sturge2024a,Montano2025}.

\begin{figure*}[htb]
    \centering
    \includegraphics[width=\linewidth]{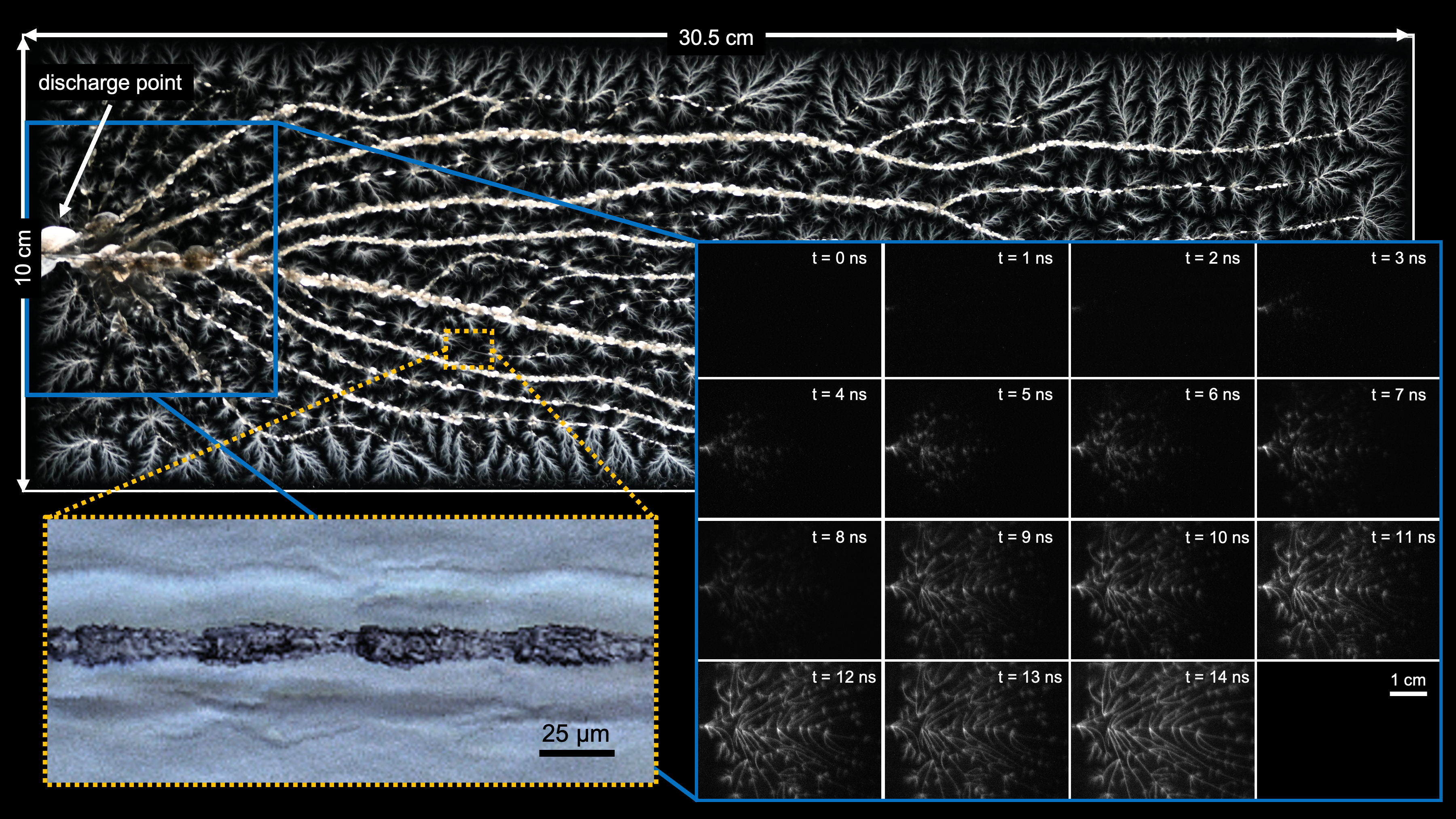}
    \caption{A backlit photograph of a discharged PMMA sample of dimensions 30.5 cm $\times$ 10 cm $\times$ 2.5 cm containing ivy-mode breakdown channels. The breakdown was initiated in the electron-charged sample at the discharge point (center left-hand side). The blue box on the photograph highlights the region that was imaged using a high-speed camera. The high-speed frames are shown in the inset outlined in blue. The interframe time for these images is -2 ns with a 3 ns exposure time, resulting in the image spacing of 1 ns. The images show the initiation of the ivy-mode breakdown, which occurs at approximately 10$^7$ m/s. The yellow dotted box on the backlit image highlights a single ivy-mode channel. The optical microscopy image below, outlined in dotted yellow, demonstrates the periodic structure of the ivy-mode channel.}
    \label{fig:LFsequence}
\end{figure*}

Of particular interest are the recently discovered ivy-mode channels \citep{Sturge2024a} (\cref{fig:LFsequence}), which represent the fastest solid-state discharge phenomenon directly observed to date. This channel type exhibits minimal branching and high-directionality. The LF propagates with discharge channels spreading radially through the PMMA while most trapped charge remains largely unperturbed post-discharge. The discharge channels form an arc known as the breakdown front, behind which breakdown has occurred and ahead of which undamaged material remains.

Many of these ivy-mode channels exhibit a striking periodic structure along their length, suggesting the presence of an underlying physical instability that drives their formation. Additionally, there is also a slight asymmetry in the periodic structure, with the peaks preferentially aligned in one direction. The periodic nature of these structures provides a unique window into the physical mechanisms governing the discharge process, as the characteristic wavelength and amplitude of these modulations contain information about the physical forces and timescales involved.

Despite the importance of dielectric breakdown to insulator applications in high-radiation environments, the physical mechanisms responsible for the periodic structures remain poorly understood. In this work, we combine materials characterization with theoretical modeling to identify the physical instability mechanism responsible for periodic structures in ivy-mode channels. Through surface imaging and spectroscopy, we investigate whether the periodic structures form during or after discharge. We then systematically evaluate three candidate instability mechanisms against our experimental observations. Our findings provide new insights into the fundamental physics of ultra-fast dielectric breakdown and establish a framework for predicting and potentially controlling discharge morphology in radiation-hardened materials, as discussed in \cref{sec:Conclusions}.

\section{Experimental methods and details}

\subsection{Charging and discharging}
\label{sec:charging_discharging}
PMMA samples underwent bombardment with a modified Varian Clinac-6 pulsed electron linear accelerator. The experiment positioned 5 cm cube PMMA samples on a rotating stage 80 cm downstream of the accelerator’s vacuum-to-air transition, a 0.08 mm-thin titanium foil window. The sample rotated approximately 1 revolution per second. At the 80 cm distance, the samples received exposure to nominally 15 nC per 3 \textmu s linac pulse with 5 MeV electrons. The pulse repetition rate was 100 Hz, with an overall exposure time of 100 seconds to achieve a total loading of 150 \textmu C -- approximately 1.5 \textmu C cm$^{-2}$ incident distributed across the four rotating irradiated surfaces. This charge was sufficient to consistently produce ivy-mode discharges during the initiated breakdown event. After the electron charging, a sharp metallic tip applied a mechanical insult to trigger breakdown, forming an LF.

To measure the current in a single channel, we analyzed isolated one-dimensional discharges from highly charged PMMA samples (1.5 \textmu C m$^{-2}$) where single-channel currents could be extracted without superposition effects from multiple channels \citep{Sturge2024a,Sturge2025}. From samples exhibiting propagation velocities above the ivy-mode threshold (1.32 $\pm$ 0.12 $\times$ 10$^6$ m/s) \citep{Sturge2024b}, we determined typical currents.

\subsection{Material analysis}

Post-discharge processing involved scoring the samples with a mini table saw at a depth of $\sim$1.6 mm, submerging them in liquid nitrogen to make the material more brittle, and then fracturing the samples in a hydraulic press. We examined the cleaved surface for exposed LF features using an optical microscope. We then cut down the 2.5 cm thick, fractured sample to a $\sim$5 mm thick, 1 $\times$ 1 cm area such that the feature of interest was centered in the final sample geometry for further characterization. Proper handling protocols prevented sample contamination.

Surface morphology analysis involved both optical imaging and electron microscopy. A Keyence VHX 7000N Digital Microscope imaged the periodic structures with white and polarized light. Post-Raman analysis, a Hitachi SU-70 Schottky field emission gun SEM characterized the surface structures. To mitigate surface charging effects from the electron gun, we coated samples with a 3 nm layer of Au/Pd with sputter deposition.

Raman mapping covered a rectangular area (75 $\times$ 148 \textmu m) along the discharge channel using a Horiba Yvon Jobin LabRam ARAMIS Confocal Raman Microscope with a 532 nm He-Ne excitation laser. We chose a step size of 4 \textmu m in the X and Y directions, corresponding to $\sim$700 pixel array. Due to strong sample fluorescence and significant surface topography, the analysis used mapping parameters of laser filter $D=0.6$, 100 \textmu m pinhole, 100 \textmu m slit, 600 diffraction grating, and $\times$10 optical lens.

\section{Results and discussion}

\subsection{Experimental}
\label{sec:Experimental}

\begin{figure*}[htb]
	\centering
	\begin{subfigure}{\linewidth}
		\centering
		\captionsetup{justification=centering,font=small}
		\includegraphics[width=\linewidth]{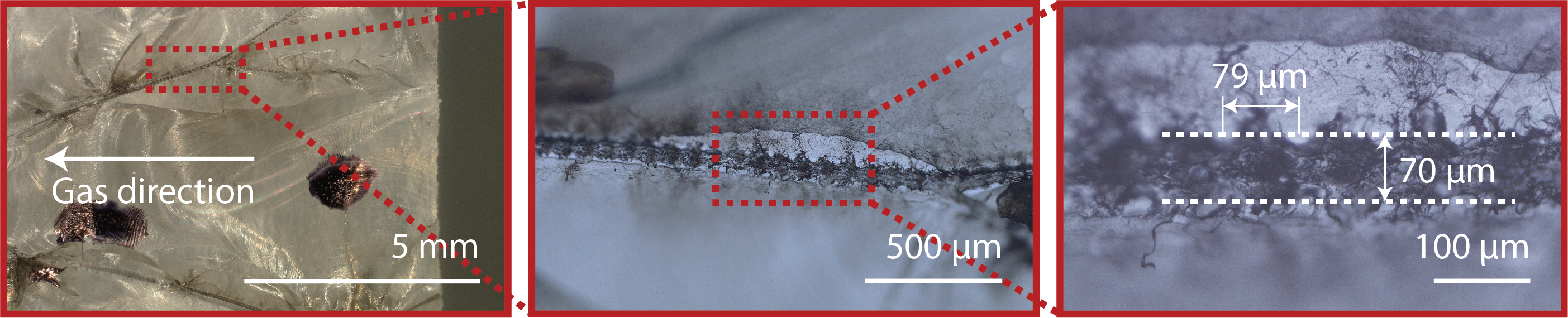}
		\caption{}
		\label{fig:optical_periodic_structures}
	\end{subfigure}
    \\
	\begin{subfigure}{\linewidth}
		\centering
		\captionsetup{justification=centering,font=small}
		\includegraphics[width=\linewidth]{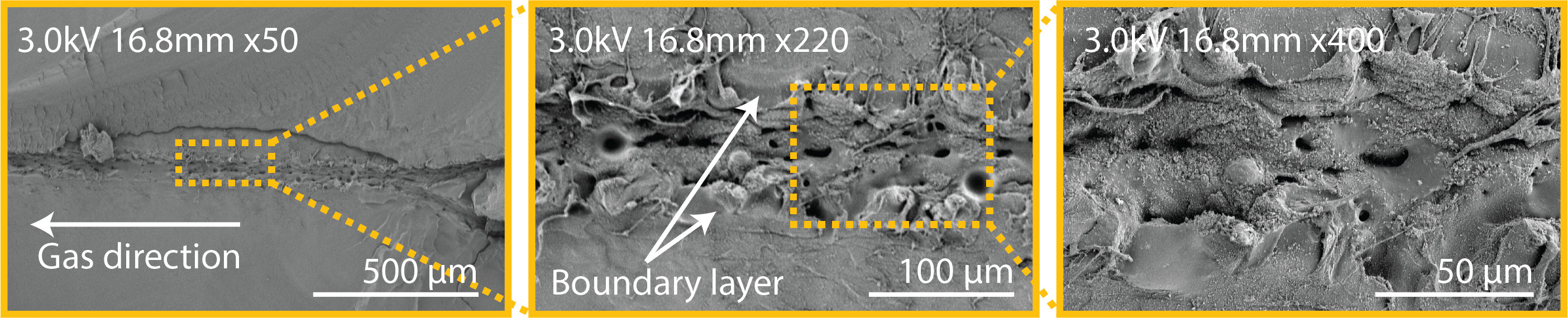}
		\caption{}
		\label{fig:sem_periodic_structures}
	\end{subfigure}
	\caption{(a) Optical microscopy under polarized light and (b) SEM images of ivy-mode periodic channels at increasing magnification. The discharge initiation point creates a pressure gradient that drives post-discharge gas flow, indicated by arrows. Boundary layers around the channels appear as slightly darker regions in SEM images, though optical visualization is limited by fracture artifacts.}
	\label{fig:channel_images}
\end{figure*}

The optical and SEM images reveal the morphological characteristics of the periodic structures (\cref{fig:channel_images}). Remarkably, these periodic structures extend many millimeters with very consistent periodicity. At increasing magnification, the channels show clear periodic width variations along their length, with boundary layers appearing as slightly darker regions in SEM images\footnote{The boundary layer in \cref{fig:optical_periodic_structures} was difficult to distinguish because the fracture pattern across the channel was not a clean break, so the boundary is distorted. \cref{fig:LFsequence} provides a clearer view of the boundary layer with an optical image.}. The SEM images clearly display enhanced surface roughness inside the channel. The periodic structures exhibit a characteristic wavelength $\lambda$ of $\sim$80 \textmu m. For a typical channel radius $R$ of $\sim$35 \textmu m, this corresponds to a dimensionless wavenumber $kR = 2 \pi R / \lambda = 2.8$.

While we observe a well-defined, dominant wavenumber, it should be noted that instability-driven periodic structures typically contain contributions from multiple wavenumbers, with the fastest-growing mode determining the primary periodicity. Our measurements and subsequent analysis focus on this dominant wavenumber, which represents the most unstable mode.

\begin{figure}
    \centering
    \includegraphics[height=0.9\textheight]{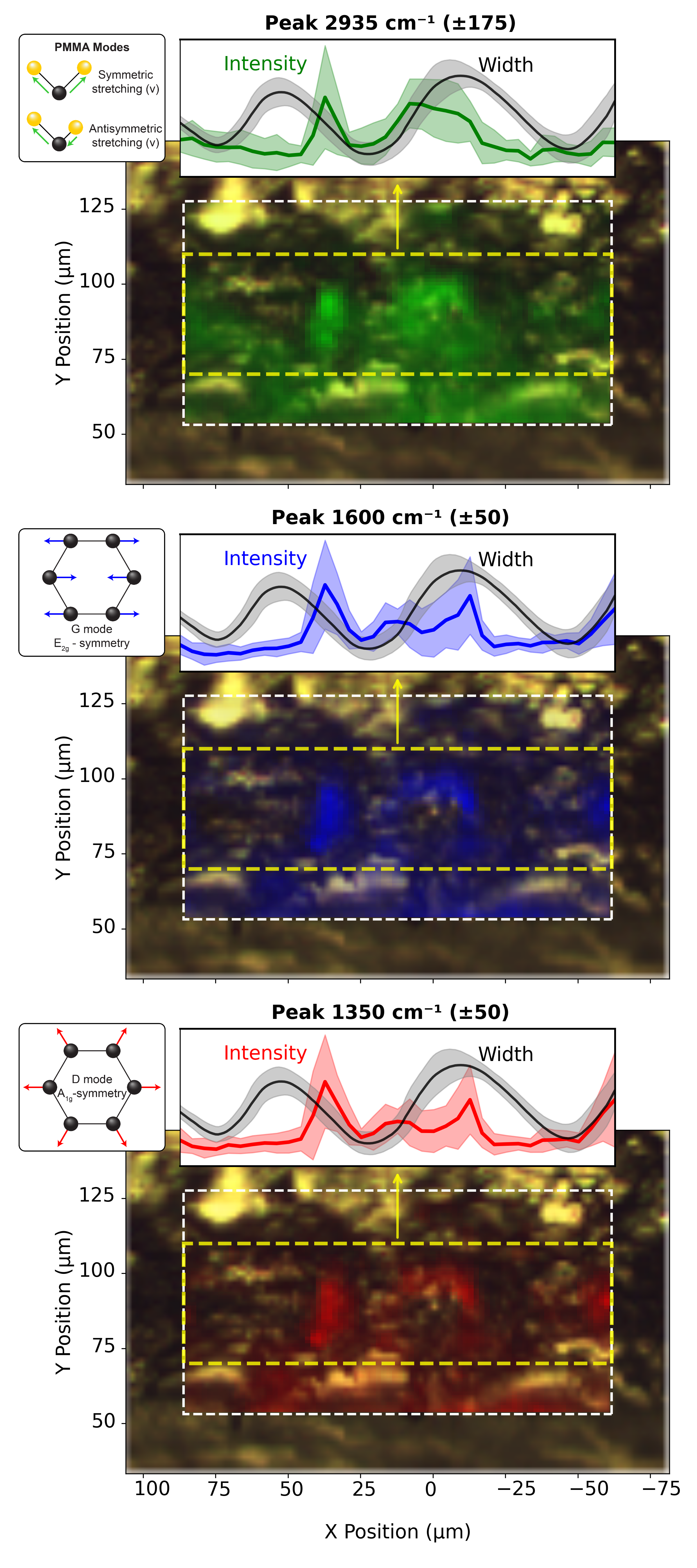}
    \caption{Raman spectral analysis of discharge channel showing D-mode (top), G-mode (middle), and PMMA stretching mode $\nu$ (bottom) intensities mapped along the channel length. The white box indicates the total scanned area, while the yellow box shows the region used for line-out analysis.}
    \label{fig:raman_analysis}
\end{figure}

To determine the material composition within these periodic structures, we performed Raman spectroscopy analysis around three characteristic wavelengths: two representing vibrational modes of amorphous carbon (D- and G-modes), and one representing PMMA ($\nu$ stretching) (\cref{fig:raman_analysis}). The presence of D- and G-peaks confirms that the breakdown channels have been carbonized, with the polymer matrix transformed into amorphous carbon during the discharge phase, and the hydrogen and oxygen exiting as gas afterwards. This carbonization occurs in all breakdown channels regardless of whether they exhibit periodic structures, though further studies are required to fully characterize the deposited material and its formation mechanisms.

The region in the white box represents the total area mapped out by the Raman scan, while the region in the yellow box represents the area in the middle of the channel that was used for a line-out. The line-out was compared to the width of the channel\footnote{The channel width along the length of the channel was difficult to measure with computer vision software because of the low resolution of the image. Two pairs of lines were hand drawn to represent the outermost and innermost possible borders of the channel, and the line plotted for the width represents the average value of channel width between these two pairs.}, and there was clear correlation between the two.

The D- and G-modes exhibit nearly identical spatial behavior along the channel length, with intensity maxima coinciding with regions of maximum channel width. The PMMA $\nu$ mode displays less pronounced peaks but maintains qualitative alignment with the channel width variations.

To quantitatively assess this spatial correspondence, we performed two one-sided t-tests (TOST) with 95\% confidence to determine the statistical equivalence between vibrational-mode periodicities and channel width. This approach tests whether the periodicities of the intensity signals and channel width are statistically equivalent within a specified tolerance. With only three observable periods (one between the two peaks and two between the three troughs) in our measurement window, TOST analysis revealed minimum equivalence margins of 32.4\%, 40.2\%, and 85.3\% for the D-, G-, and PMMA modes, respectively ($p < 0.05$), relative to the measured channel width periodicity of 60.1 $\pm$ 6.1 \textmu m. The substantially smaller margins required for the D- and G-modes compared to the PMMA mode provide strong quantitative evidence of spatial correlation between carbon-related vibrational modes and channel geometry.

Since carbon deposition requires plasma conditions, and carbon flux to channel walls increases with plasma temperature \citep{Konuma1992,Yarin2006}, this correlation between the Raman signal and channel width demonstrates that plasma temperatures were higher in the wider channel regions. This temperature variation during the plasma discharge phase will be explored further in \cref{sec:entropy_mode} when evaluating candidate instability mechanisms. Although we cannot directly measure plasma temperature during the nanosecond discharge, the carbon deposition pattern provides evidence that periodic structure formation coincides with the discharge phase rather than occurring during any subsequent phase. 

\subsection{Theoretical modeling of instability growth}
\begin{figure*}[htb]
    \centering
    \begin{subfigure}{0.32\linewidth}
        \centering
        \captionsetup{justification=centering,font=small}
        \includegraphics[width=\linewidth]{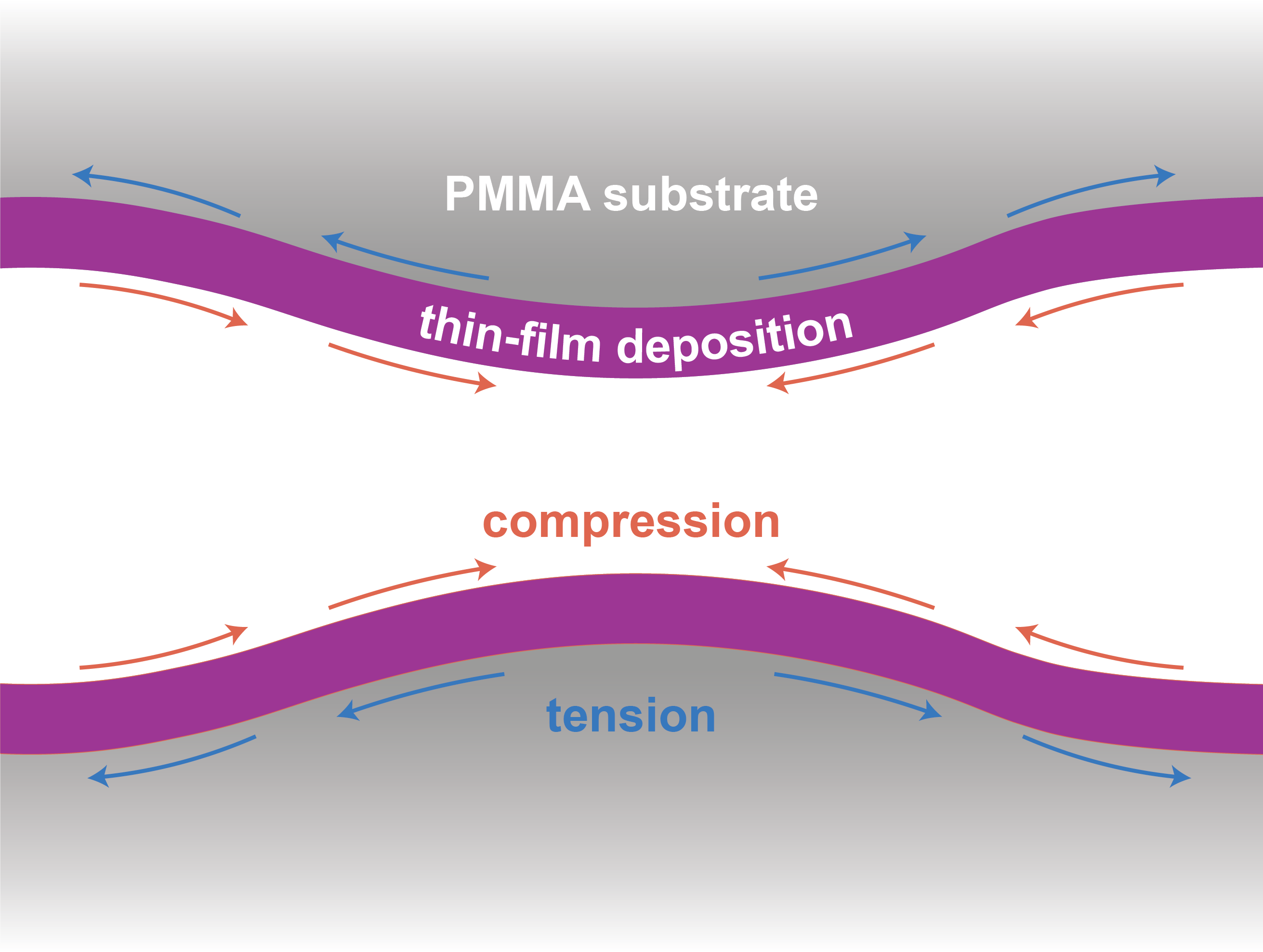}
        \caption{}
        \label{fig:atg_instability}
    \end{subfigure}
    \begin{subfigure}{0.32\linewidth}
        \centering
        \captionsetup{justification=centering,font=small}
        \includegraphics[width=\linewidth]{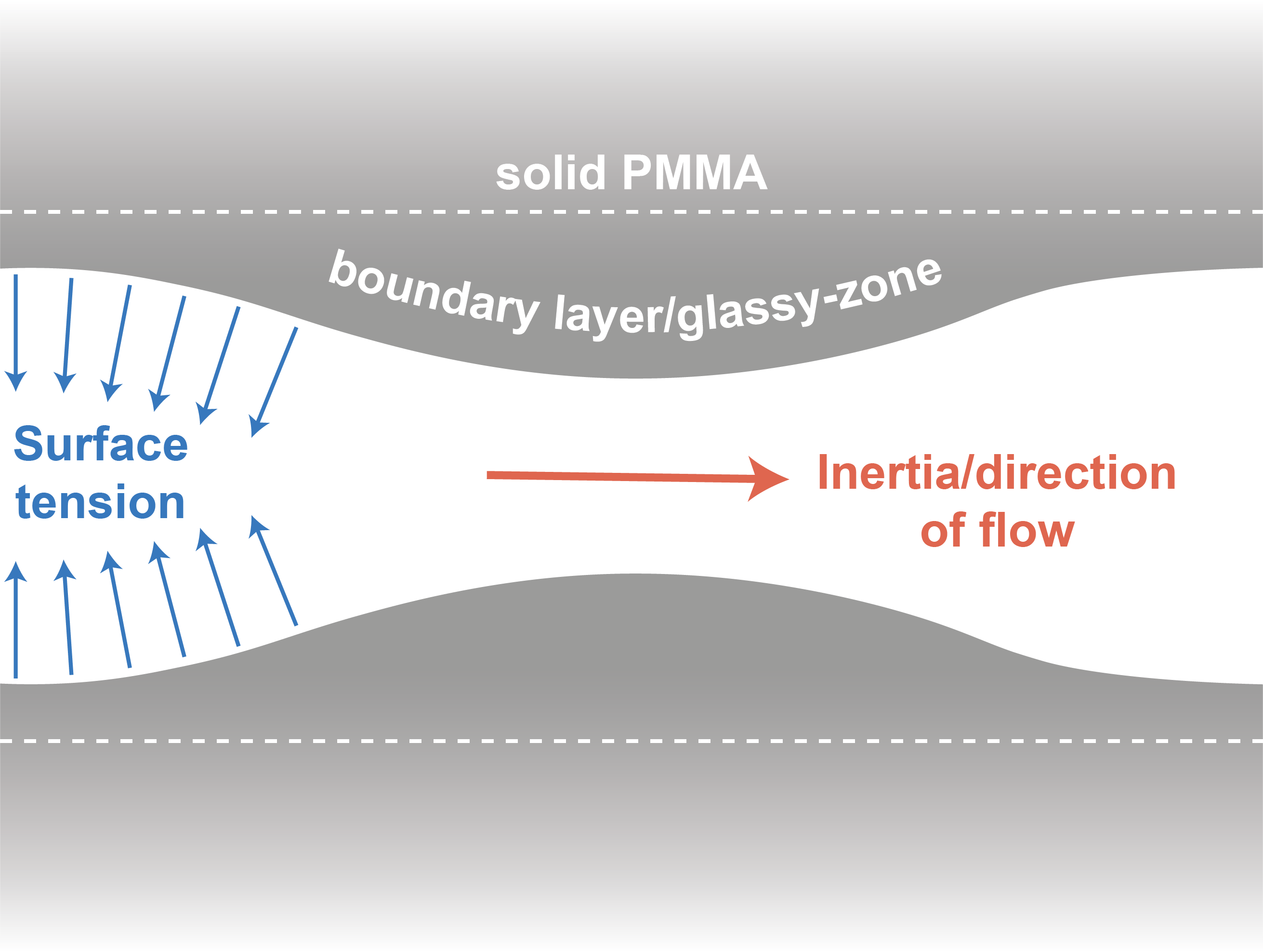}
        \caption{}
        \label{fig:pri_instability}
    \end{subfigure}
    \begin{subfigure}{0.32\linewidth}
        \centering
        \captionsetup{justification=centering,font=small}
        \includegraphics[width=\linewidth]{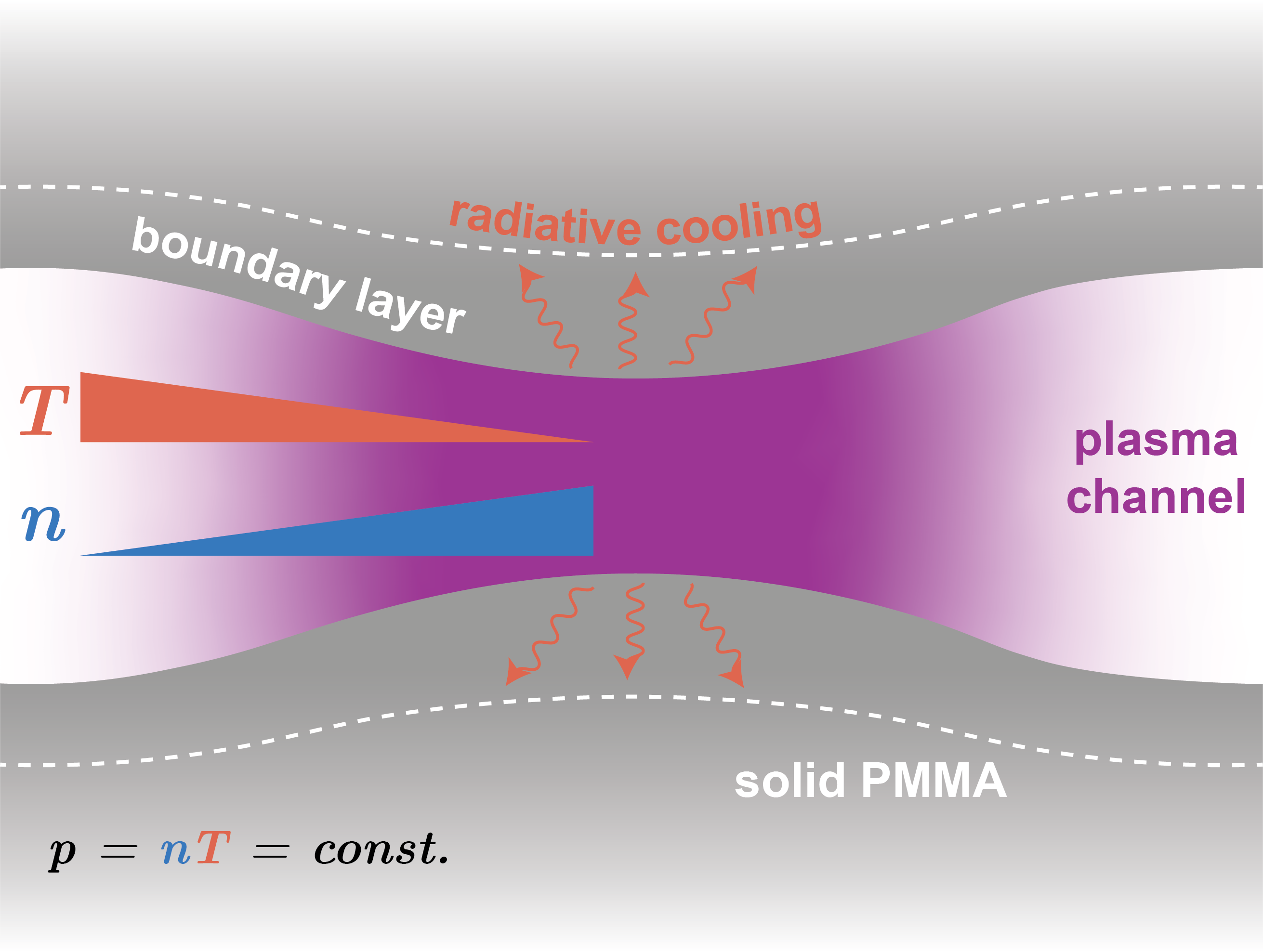}
        \caption{}
        \label{fig:entropy_mode}
    \end{subfigure}
    \caption{Schematic diagrams of three candidate instability mechanisms for periodic channel structures. (a) Asaro-Tiller-Grinfeld (ATG) instability: elastic stress relief through surface undulations in bi-material systems with mismatched thermal expansion coefficients. (b) Plateau-Rayleigh instability (PRI): surface tension-driven breakup of cylindrical fluid jets or cavities. (c) Z-pinch entropy mode: temperature-gradient-driven instability in high current density plasma discharges.}
    \label{fig:instability_mechanisms}
\end{figure*}

The periodic structures observed in the PMMA channels suggest an underlying physical instability mechanism. We systematically evaluated three candidate instabilities based on their predicted growth rates and physical relevance to the experimental conditions: the Asaro-Tiller-Grinfeld (ATG) instability, the Plateau-Rayleigh instability (PRI), and the z-pinch entropy mode. The instabilities operate on different time scales and involve different competing physical forces, as illustrated (\cref{fig:instability_mechanisms}) and described in \cref{tab:instabilities}. The following sections describe the instabilities in detail, carefully considering whether each model fits the observed behavior.

\begin{table*}
	\begin{center}
    \caption{Comparison of three candidate instability mechanisms for explaining periodic channel structures.}
    \begin{tblr}{colspec=lccc}
        \hline \hline
        & ATG & PRI & Z-Pinch \\
        \hline
        Diagram & \cref{fig:atg_instability} & \cref{fig:pri_instability} & \cref{fig:entropy_mode} \\
        {Competing\\factors} & {Elasticity vs. surface\\tension} & {Inertia vs. surface\\tension} & {Density vs. temperature\\exchange} \\
        Timescale & s & ms & ns \\
        Directionality? & No & Yes & Yes \\
        \hline \hline
    \end{tblr}
    \label{tab:instabilities}
    \end{center}
\end{table*}

\subsubsection{Asaro-Tiller-Grinfeld instability}
We first consider the ATG instability, which would occur during the post-discharge cooling phase when the system approaches thermal equilibrium. Based on the Raman map in \cref{sec:Experimental}, the plasma discharge deposits a thin carbon layer on the inner channel walls, creating a bi-material system with the PMMA substrate. As the system cools, the different thermal expansion coefficients of the carbon film and PMMA generate interfacial stresses that can be relieved through periodic surface undulations \citep{Grinfeld1993,Grinfeld1993-2,Kohlert2003,Colin1997,Colin1999} (\cref{fig:atg_instability}).

\begin{figure}[htb]
    \centering
    \includegraphics[width=0.9 \linewidth]{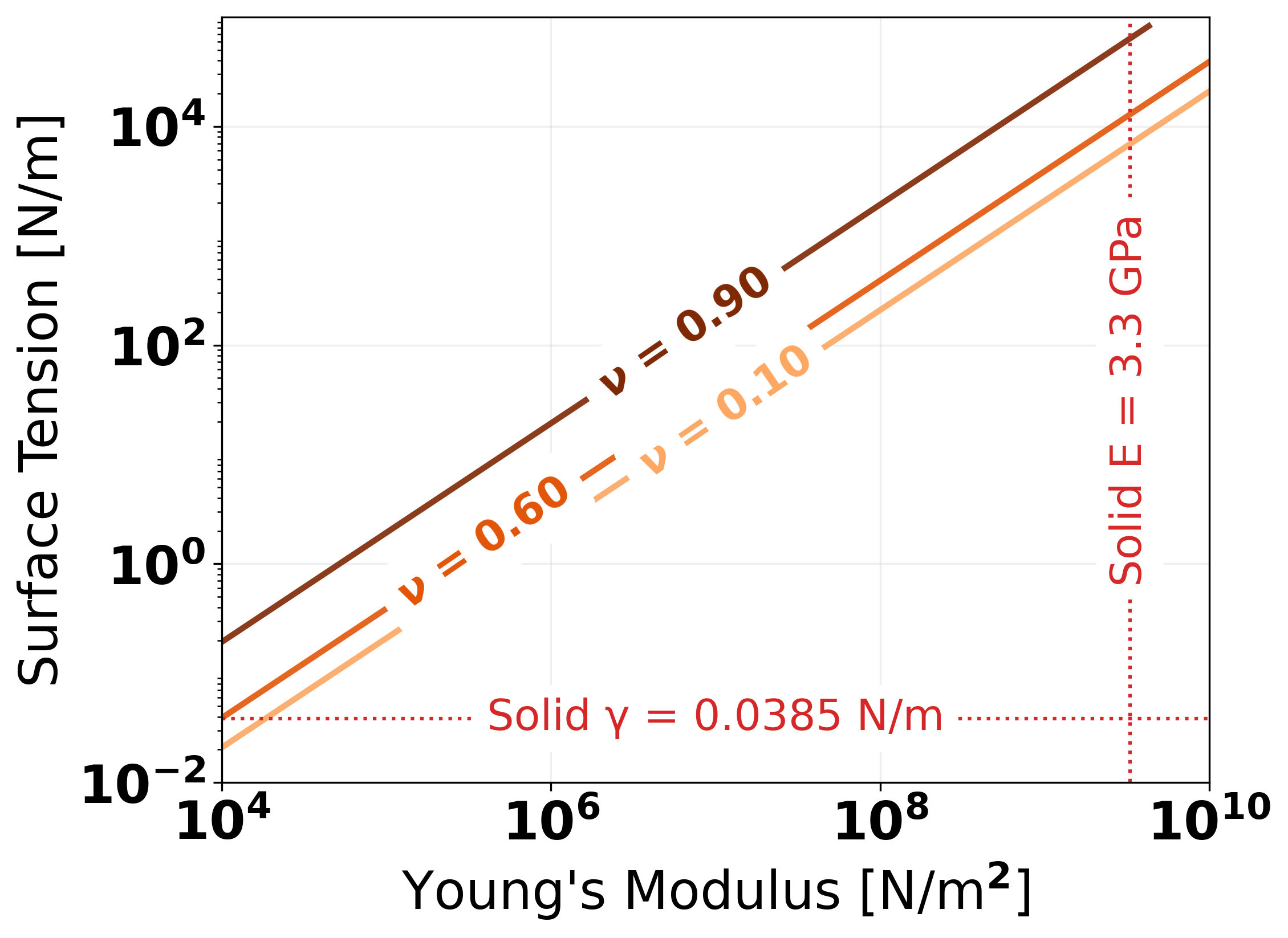}
    \caption{Required elastic modulus as a function of Poisson's ratio for the ATG instability to match observed wave numbers. The horizontal and vertical lines show accepted values in solid PMMA for the surface tension and elastic modulus, respectively.}
    \label{fig:material_analysis}
\end{figure}

We extended previous analyses of cylindrical geometries \citep{Cammarata1994,Colin1997,Colin1999} and adapted the model for a cylindrical wire from \citet{Huang2024} to analyze a cylindrical cavity under thermal stress. The governing force balance equations are
\begin{equation}
\label{eq:force_balance_1}
\begin{split}
    0 &= A \left\lbrace k (1 - 2 \nu) J_0(kR) - \frac{1}{R} \left[ 1 - k^2 R (1 + \nu) \frac{\gamma}{E} \right] J_1(k R) \right\rbrace\\
    &+ B \left\lbrace k \left[ 1 - 2 \nu + k^2 R (1 + \nu) \frac{\gamma}{E} \right] J_0(k R).- (1 - 2 \nu) k^2 R J_1(k R) \right\rbrace,
\end{split}
\end{equation}
\begin{equation}
\label{eq:force_balance_2}
\begin{split}
    0 &= A (1 - 2 \nu) J_1(k R)\\
    &+ B \left[ 2 (1 - \nu) J_1(k R) - (3 - 2 \nu) k R J_0(k R) \right],
\end{split}
\end{equation}
where $A$ and $B$ are constants determined by boundary conditions, $k$ is the axial wave number, $R$ is the average radius of the channel, $\nu$ is Poisson's ratio, $J_0$ and $J_1$ are the zeroth- and first-order Bessel functions of the first kind, respectively, $\gamma$ is the surface tension, and $E$ is the modulus of elasticity. Setting the determinant of these equations to zero yields the dispersion relation
\begin{equation}
\begin{split}
    0 &= J_0^2(kr) k^2 R (1 - 2 \nu)(3 - 2 \nu) \\
    &+ J_1^2(kr) \left[ 2 k^2 (1 - \nu^2) \frac{\gamma}{E} - \frac{2}{R} (1 - \nu) + k^2 R (1 - 2 \nu) \right]\\
    &+ 4 J_0(kr) J_1(kr) k \left[ 1 - \nu - k^2 R \frac{\gamma}{E} (1 - \nu^2) \right].
\end{split}
\end{equation}
Rather than solving for $k$ directly, we determined what combination of material parameters would satisfy the observed wave number and channel radius ($kR = 2.8$, from \cref{fig:optical_periodic_structures}). The results, shown in \cref{fig:material_analysis}, reveal a fundamental inconsistency with known material properties.

For typical PMMA surface tensions of 0.0385 N/m \citep{Kwok1998}, matching the observed $k$ and $R$ values would require elastic moduli of $\sim$10$^3$-10$^4$ Pa across the range $\nu = 0.3-0.9$. This is many orders of magnitude below the accepted elastic modulus of PMMA ($\sim$3.3 GPa \citep{Abdel-Wahab2017}). Therefore, the ATG instability cannot explain the observed periodic structures.

\subsubsection{Plateau-Rayleigh instability}
During the post-discharge phase, evaporated material flows back through the channels as neutral gas. The intense UV radiation from the plasma creates a boundary layer of heated PMMA that exhibits viscoelastic \citep{McLoughlin1952,Dooling1998} or even semi-liquid \citep{Mathiesen2014} behavior. Under these conditions, the Plateau-Rayleigh instability could potentially develop.

The PRI occurs when surface tension forces overcome fluid inertia, causing cylindrical jets to develop periodic undulations that can eventually break into droplets (\cref{fig:pri_instability}). In viscoelastic materials, the competing forces are elastic energy and capillary energy rather than inertia and surface tension \citep{Tamim2021}.

We applied the viscoelastic PRI model of \citet{Tamim2021}, which provides stability criteria for cylindrical cavities. In the purely fluid limit (no elastic effects), the maximum stable dimensionless wave number is $kR = 1$. Viscoelastic effects only serve to further stabilize the system against PRI.

Our measurements consistently showed $kR = 2.8 > 1$ along the channel length (\cref{fig:channel_images}), placing the observed geometry well within the stable regime for PRI. Therefore, the Plateau-Rayleigh instability cannot account for the periodic structures, particularly when viscoelastic stabilization is considered.

\subsubsection{Z-pinch entropy mode}
\label{sec:entropy_mode}
The most promising mechanism occurs during the plasma discharge itself. The high current discharge creates a dense plasma column that behaves as a z-pinch configuration \citep{Haines2011}. While the classical "sausage" instability produces $m=0$ modes through non-uniform magnetic pinching, it predicts the fastest growth at infinitesimally small wavelengths ($k \to \infty$) under ideal magnetohydrodynamic conditions \citep{Coppins1988,Liberman1999}. Since we observe finite, well-defined wavelengths, this mechanism is unlikely.

Instead, we propose that the entropy mode \citep{Kadomtsev1960,Ware1962,MIKHAILOVSKAIA1968,MIKHAlLOVSKII1971,Kesner2000,Kesner2002,Ricci2006} drives the instability. This mode becomes dominant when $k\rho_i \gtrsim 1$, where $\rho_i$ is the ion gyroradius \citep{Angus2019}. In this regime, ions cannot conserve their magnetic moment, leading to non-adiabatic heating and temperature-gradient-driven instabilities. The entropy mode grows by coupling misaligned density and temperature gradients while maintaining constant pressure, creating periodic structures through the growth of these gradients.

The z-pinch entropy mode provides the best explanation for our experimental observations. Our Raman analysis (\cref{fig:raman_analysis}) demonstrates that carbon deposition is directly correlated with channel width, establishing two critical facts: (1) that periodic structures formed during the discharge phase when carbon formation occurred, and (2) that plasma temperatures were higher in the wider channel regions. This temperature gradient is precisely what drives the entropy mode instability, in which the plasma temperature is highest in the widest part of the discharge channel (\cref{fig:entropy_mode}) -- exactly matching our experimental observations.

\begin{figure}[htb]
    \centering
    \includegraphics[width=0.9 \linewidth]{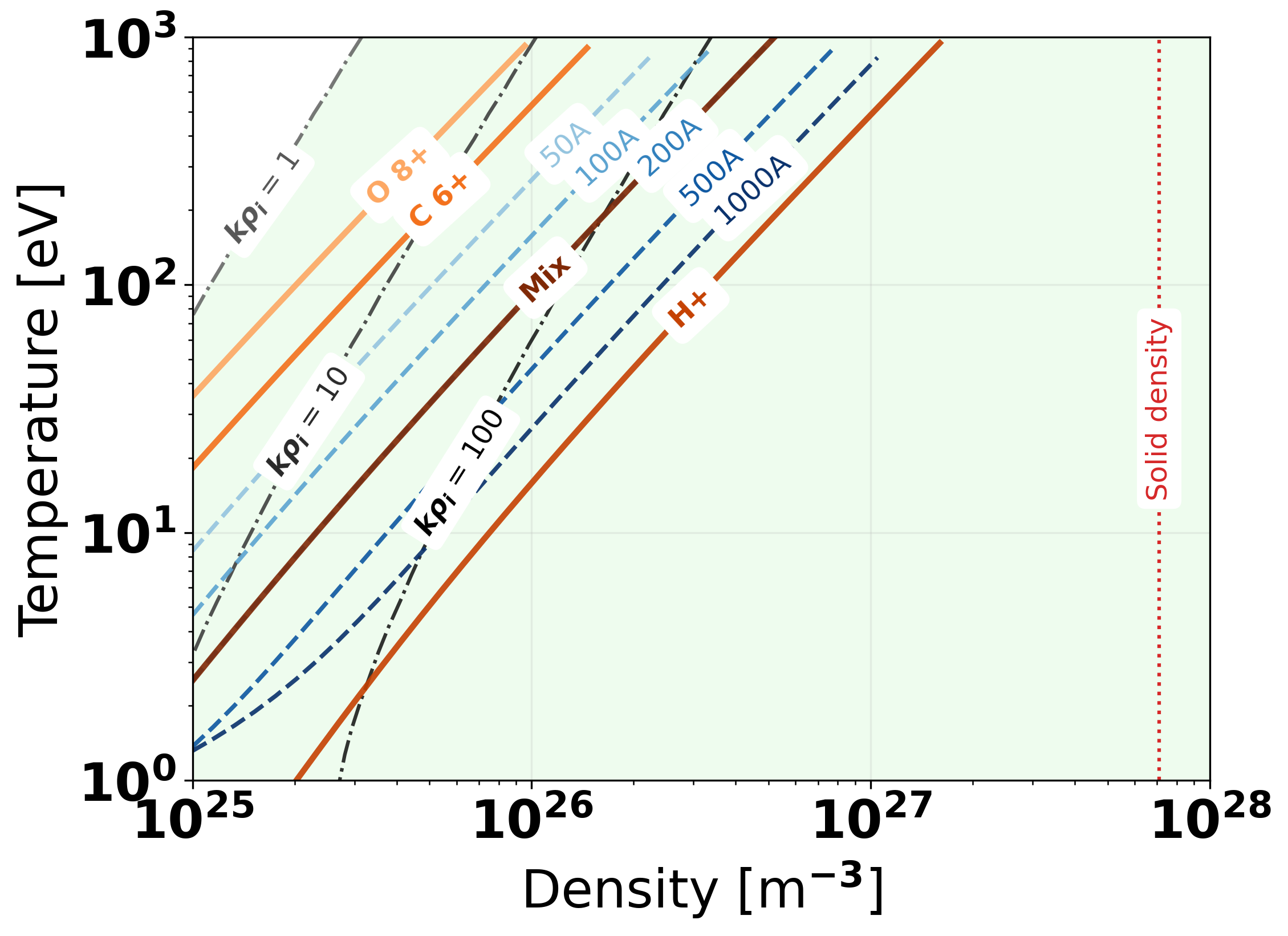}
    \caption{Predicted plasma parameters for z-pinch entropy mode instability matching observed wave numbers. Solid lines show constant current (200 A) with varying ion composition. Dashed lines show constant composition of PMMA (1/3 C$^{6+}$, 2/15 O$^{8+}$, 8/15 H$^+$) with varying current from 100--1000 A. Grey dash-dot lines indicate $k \rho_i$ contours. The green shaded region is where the entropy mode is valid, i.e. $k \rho_i > 1$}
    \label{fig:combined_analysis_ion_focus}
\end{figure}

We modeled the entropy mode using the dispersion relation (Eq. 40) from \citet{Angus2019}, varying the plasma density and temperature to match the observed wave number. \cref{fig:combined_analysis_ion_focus} shows the results for two scenarios: constant current (200 A) with varying ion composition (solid lines), and constant composition (1/3 C$^{6+}$, 2/15 O$^{8+}$, 8/15 H$^+$ representing the chemical formula for PMMA, C$_5$O$_2$H$_8$) with varying current (dashed lines). The gray dashed-dotted lines represent countours of $k \rho_i$, confirming that most solutions lie in the entropy mode regime, $k \rho_i > 1$.

The predicted plasma parameters are physically reasonable: electron densities of 0.1--1\% of solid density, electron temperatures of 10--100 eV, and individual channel currents of 100--500 A. This was validated by current measurements as described in \cref{sec:charging_discharging}, for which the average peak current was 217 $\pm$ 21 A per channel. This experimental value falls well within our theoretical prediction range, providing additional support for the entropy mode mechanism.

Although direct measurements of plasma density, temperature, and composition remain challenging, the consistency between theoretical predictions and experimental observations -- including both morphological features and current measurements -- provides strong evidence that the z-pinch entropy mode is the dominant instability mechanism responsible for the periodic channel structures.

While our analysis identifies the z-pinch entropy mode as the governing mechanism, periodic structures do not appear in all breakdown channels, nor even in all ivy-mode channels. We hypothesize that smaller-radius channels combined with higher current in ivy-mode channels have higher current densities that exceed some instability threshold (perhaps $k \rho_i > 1$), while larger-radius channels or those with lower current operate below this threshold and exhibit discharge without periodicity. The mechanism governing radius selection of an individual channel during breakdown remains unclear and represents an important direction for future investigation.

\subsection{Phenomenological timeline}
\begin{figure*}[htb]
    \centering
    \includegraphics[width=0.8 \linewidth]{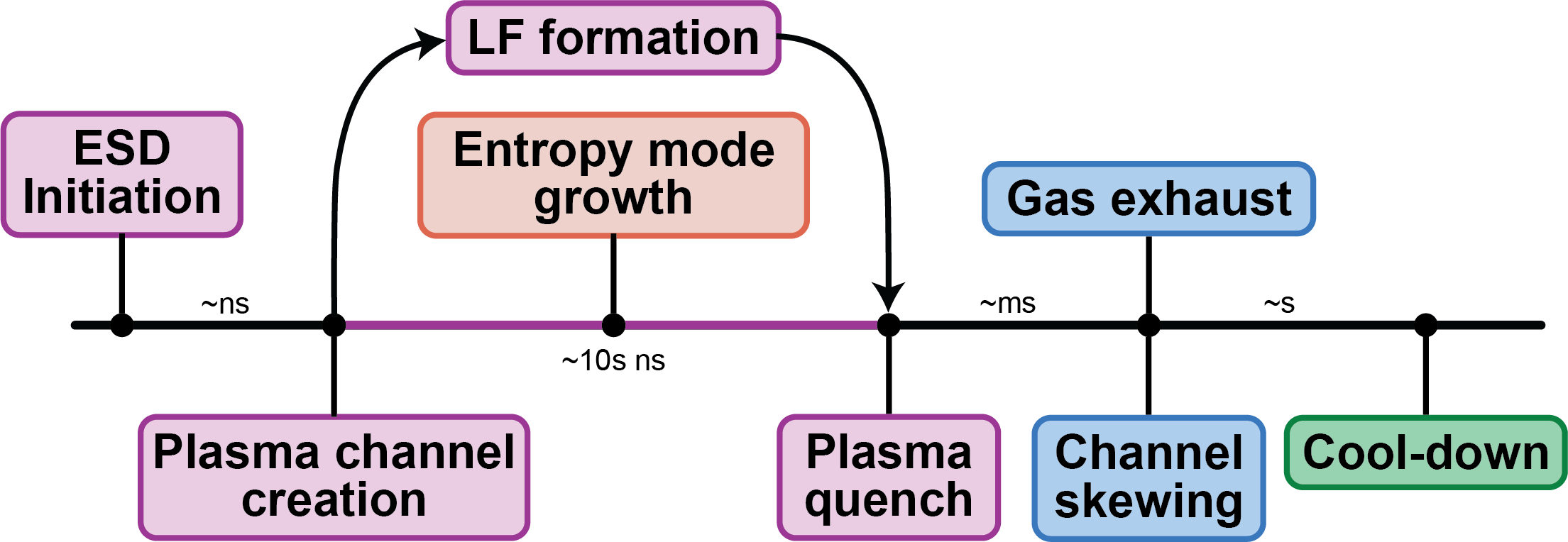}
    \caption{Timeline describing the key events during the formation of the periodic structures. The plasma processes are highlighted in purple, the instability in orange, the gas dynamics in blue, and the thermal cool-down in green. The approximate time between each process is provided along the timeline. The arrows to and from `LF formation' indicate the beginning and end of that process.}
    \label{fig:pmma_timeline}
\end{figure*}

Our theory posits that the formation of periodic structures in ivy-mode channels involves a sequence of processes occurring across multiple timescales (\cref{fig:pmma_timeline}). Following breakdown initiation, plasma channel creation occurs within nanoseconds. During this plasma phase, a high-current discharge creates z-pinch-like conditions.

The entropy mode instability drives periodic structure formation during the $\sim$tens of nanoseconds of plasma discharge itself. Furthermore, radiation from the plasma creates a boundary region of hot, glassy material in the channel walls. In support of this mechanism, \cref{fig:LFsequence} shows that the boundary region itself is \textit{also} periodic, providing direct evidence that the instability occurs during the plasma discharge phase when radiative heating creates the boundary region.

After plasma quench ($\sim$milliseconds), gas exhaust occurs as evaporated material flows through the channels toward the discharge point. As the glassy boundary layer has not yet had sufficient time to cool, this flowing gas causes skewing of the periodic structures, explaining the observed phase shifts in D- and G-mode Raman peaks. Finally, thermal cooling over several seconds returns the material to ambient conditions, while preserving the periodic structures created during the plasma phase.
 
\section{Conclusions}
\label{sec:Conclusions}
Through combined materials characterization and theoretical modeling, we have identified the z-pinch entropy mode as the physical mechanism responsible for periodic structures in dielectric breakdown of PMMA. Optical and SEM images indicate that the boundary layer is also periodic, and Raman spectroscopy revealed that carbon deposition correlates with channel width variations. These observations demonstrate that (1) these structures form during the plasma discharge phase, and (2) the plasma temperature is higher in the wider channel regions. The entropy mode operates on nanosecond timescales consistent with plasma formation and produces wavelengths that match experimental observations when plasma parameters reach physically reasonable values. Discharge current measurements from isolated breakdown channels validate theoretical predictions, providing strong evidence for this plasma instability mechanism.

These findings advance our fundamental understanding of breakdown physics in insulators and establish a framework for predicting discharge morphology. The plasma in these channels probably had an electron temperature of 10--100 eV and an electron density of 0.1--1\% of solid density. By identifying the mechanism responsible for the growth of these periodic structures, this work opens new avenues for potentially controlling breakdown patterns, with direct implications for improving the reliability of insulators in high-radiation environments.

Future work should further characterize the carbon material deposited in the channels and its formation. Additional work should investigate the relationship between breakdown conditions and channel radius selection, which appears to govern whether periodic structures develop through the entropy mode instability.

\section*{CRediT authorship contribution statement}

\textbf{Nick R. Schwartz}: Conceptualization, Data curation, Formal analysis, Investigation, Methodology, Software, Validation, Visualization, Writing – original draft, Writing – review and editing.
\textbf{Bryson C. Clifford}: Conceptualization, Formal analysis, Investigation, Methodology, Validation, Visualization, Writing – original draft, Writing – review and editing.
\textbf{Carolyn Chun}: Supervision, Writing – original draft, Writing – review and editing.
\textbf{Emily H. Frashure}: Investigation.
\textbf{Kathryn M. Sturge}: Investigation, Validation, Visualization, Writing – original draft, Writing – review and editing.
\textbf{Noah Hoppis}: Conceptualization, Data curation, Writing – review and editing.
\textbf{Holly Wilson}: Investigation, Project administration.
\textbf{Meryl Wiratmo}: Investigation, Writing – review and editing.
\textbf{Jack R. FitzGibbon}: Investigation.
\textbf{Ethan T. Basinger}: Investigation.
\textbf{Brian L. Beaudoin}: Supervision.
\textbf{Raymond J. Phaneuf}: Conceptualization, Writing – review and editing.
\textbf{John Cumings}: Conceptualization, Methodology, Resources, Supervision, Writing – review and editing.
\textbf{Timothy W. Koeth}: Conceptualization, Funding acquisition, Methodology, Project administration, Resources, Supervision, Writing – original draft, Writing – review and editing.

\section*{Declaration of competing interest}

The authors declare that they have no known competing financial interests or personal relationships that could have appeared to influence the work reported in this paper.

\section*{Acknowledgments}

This research was developed with funding from the Defense Advanced Research Projects Agency (DARPA). The views, opinions and/or findings expressed are those of the authors and should not be interpreted as representing the official views or policies of the Department of Defense or the U.S. Government.

\section*{Data availability}

Data will be made available on request.

\bibliographystyle{elsarticle-num-names} 
\bibliography{references}

\end{document}